\documentclass{amsart}
 \usepackage[dvipdf]{graphicx}
\pdfoutput=1 
\usepackage{epsf}
\usepackage{color}

\usepackage[sort,numbers]{natbib}

\begin{document}
\title[Co-detection of acoustic emissions]{Co-detection of acoustic emissions during failure of heterogeneous media: new perspectives for natural hazard early warning}

\author{J.~Faillettaz}
\address{3G, University of Zurich, 8057 Zurich, Switzerland}

\author{D.~Or} 
\address{STEP, ETH Z\"urich, 8092 Z\"urich, Switzerland}

\author{I. Reiweger}
\address{WSL, SLF, Davos, Switzerland}

\markleft{J. FAILLETTAZ ET AL.}

\begin{abstract}
A promising method for real time early warning of gravity driven rupture that considers both the heterogeneity of natural media and characteristics of acoustic emissions attenuation is proposed. The  method capitalizes on co-detection of elastic waves emanating from micro-cracks by multiple and spatially separated sensors. Event co-detection is considered as surrogate for large event size with more frequent co-detected events marking imminence of catastrophic failure.
Using a spatially explicit fiber bundle numerical model with spatially correlated mechanical strength and two load redistribution rules, we constructed a range of mechanical failure scenarios and associated failure events (mapped into AE) in space and time. Analysis considering hypothetical arrays of sensors and consideration of signal attenuation demonstrate the potential of the co-detection principles even for insensitive sensors to provide early warning for imminent global failure.
\end{abstract}

\maketitle

\section{Introduction}

Gravity driven instabilities in natural earth materials including rockfalls, landslides, snow avalanches or glacier break-offs
represent a important class of natural hazards in mountainous regions. Reliable prediction of imminence of such failure events combined
with timely evacuation remain a challenge due to the nonlinear nature of geological material failure hampered by inherent heterogeneity, unknown initial mechanical state, and complex load application (rainfall, temperature, etc.) that hinder predictability. 
 Nevertheless, such materials exhibit certain characteristics during failure, they break gradually with the weakest parts
breaking first, during which they produce precursory ”micro-cracks” and associated elastic waves traveling in the material. The
monitoring of such acoustic/micro-seismic activity offers valuable information concerning the progression of damage and
imminence of global failure  (\cite{Michlmayr&al2012}). Acoustic emission methods are advancing rapidly and are expected to provide new insights into the imminence of instabilities and in some cases it has been applied to natural gravity-driven instabilities such as cliff collapse \citep{Amitrano&al2005}, slope instabilities \citep{Dixon&al2003,Kolesnikov&al2003,Dixon&Spriggs2007} or failure in snow pack \citep{vanHerwijnen&Schweizer2011,Reiweger&al2015}.  

An important technical challenge hindering application of such early warning methods is the attenuation of elastic waves propagating through the natural media, thereby introducing ambiguity in the interpretation of the magnitude (severity), or leading to loss of detection for events away from the sensor. Hence, a micro-crack event would be measured as a large event if occurring close to the sensor, and as a small event if far from the sensor (or may not be detected at all). A more complete picture of acoustic emissions or micro- seismic activity requires deployment of a dense network of sensors that enables localization of sources and thus the determination of initial energy released with each event. However, such networks are prohibitively costly, and difficult to analyze in real time over practical scales of interest. Is it possible to directly analyze in real time the measured micro-seismic activity to infer the slope mechanical status? The objective of this work is to proposed a method focusing on event co-detection for identifying large failure events and use of their frequent occurrence as precursors for imminent failure.   

The paper is organized as follows:
We first provide a qualitative description of the observation problem and the processes leading to attenuation. Taking advantage of both the heterogeneity of natural media and the attenuation phenomenon, a new method based on co-detection properties is introduced.
A quantitative analysis is then performed using a numerical model based on the classical Fiber Bundle Model. Introducing a basic attenuation law in such simple models enables to directly compare un-attenuated and attenuated acoustic activity (and also avalanche size-frequency distribution) at any location. After investigating effects of attenuation phenomena on possible stability assessment, the co-detection method is examined. Finally, the potential application of this method to early warning purposes are discussed.

\section{Qualitative approach: The observation problem}

\subsection{Effects of elastic wave attenuation}

Natural slopes act as a lowpass filter. 
Acoustic waves propagating in natural media attenuate due to geometrical reduction in elastic energy density with distance from the source (geometric spreading), e.g., the energy of spherical wavefront emanating from a point source is distributed over a spherical surface of ever increasing size. Since energy is proportional to amplitude squared, an inverse square law for energy translates to a $ 1/r$ decay law for amplitude.

\begin{figure*}
\noindent\includegraphics[width=\textwidth]{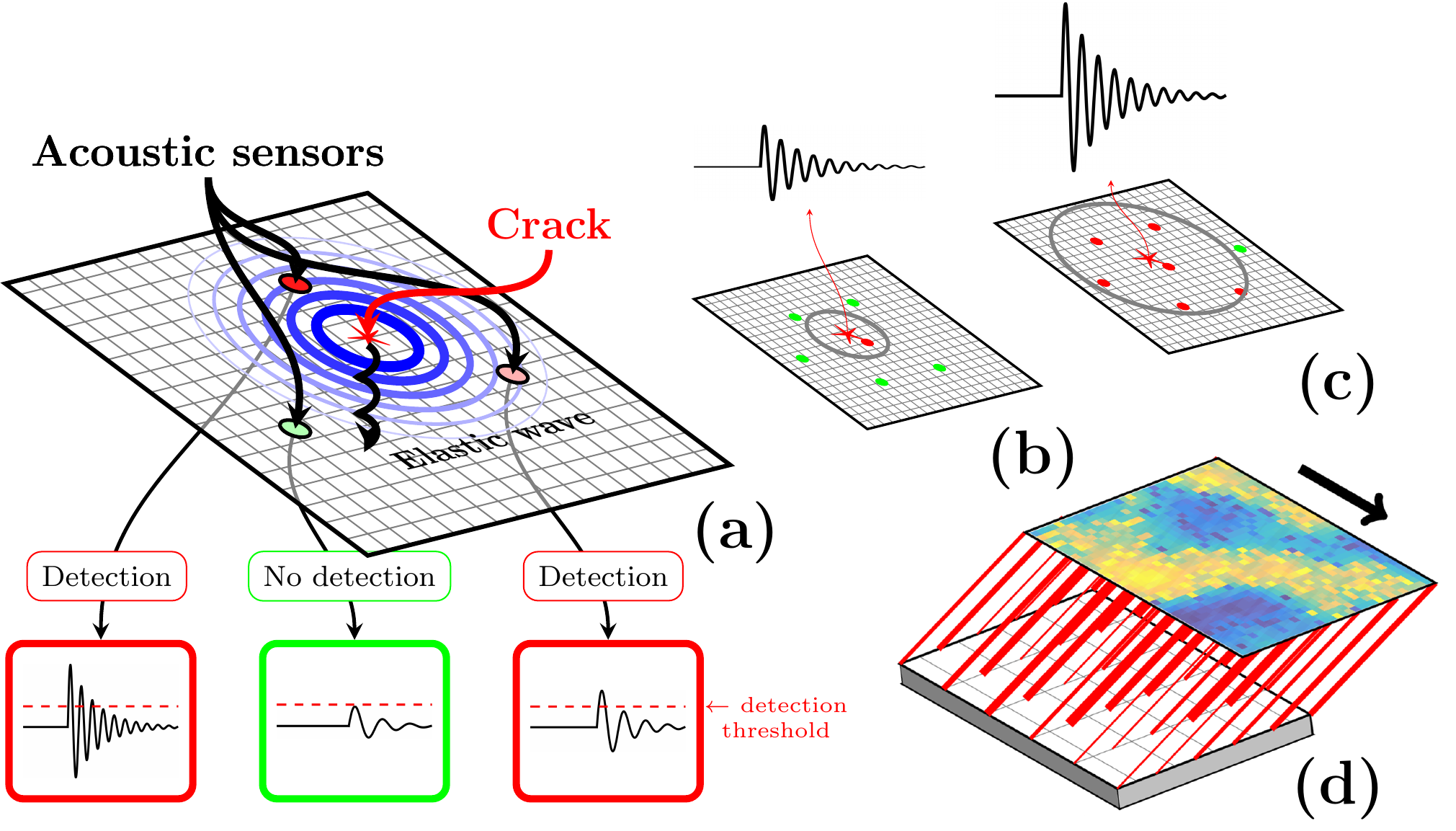}
\caption{\label{obs_attenuation}\textit{(a)}: Schematic view of the attenuation problem. Star indicates the initial crack location and blue circles the resulting elastic wave that propagates through the medium. Schematic wave forms recorded at the different sensors are also represented. \textit{(b), (c)}: Schematic illustration of co-detection method with on sensor network for a small (\textit{b}) and a large event (\textit{c}): Star indicates micro-crack location with its associated wave form on top, colored disks indicate sensor location. Grey circles represent distance of detection of the event knowing detection threshold of sensors. Dark red disks shows the sensors that detected the signal, light green the sensors that do not detect the event. The larger the event, the higher number of sensors detect the event simultaneously.\textit{(d)}: Schematic view of the 2D spatially explicit Fiber Bundle Model as an analogous to a natural slope. Width of red lines indicates fiber strength, also represented on the top (from weak dark blue to strong light yellow).}
\end{figure*}

Additionally, changes in material properties and heterogeneities give rise to signal scattering, excitation of other modes of wave propagation and conversion of elastic energy to heat. 
The resulting Acoustic Emission (AE) signal amplitude decay and dispersive effects are enhanced in the presence of pore water and air-water interfaces (especially in the presence of pore water) \citep{Oelze&al2002,Wang&Santamarina2007,Muller&al2010}. Consequently, attenuation phenomena are enhanced in wet materials close to rainfall induced failure.
Attenuation and scattering effects are frequency dependent, higher frequencies of the waveform attenuate preferentially. 
Attenuation of propagating acoustic signals (elastic waves) modifies an event (i.e. a crack in the material) such that it may be observed and recorded differently by an acoustic sensor depending on its location (Fig. \ref{obs_attenuation}).

Acoustic sensors detect a signal if the wavefront amplitude exceeds a certain detection threshold (Fig. \ref{obs_attenuation}). Thus the detection range for a given sensor location depends on both the sensor detection threshold and on AE signal initial amplitude.
Additionally, an  event is detected if the "Signal-to-noise ratio" is sufficiently large to distinguish it from background noise.

Attenuation phenomenon introduces ambiguity into the interpretation of the signal amplitude (distance and attenuation mask signal size).
This dependency plays an important role in the resulting amplitude-frequency distribution derived from sensor measurements, since exchanges of amplitude from the real to the attenuated - i.e., the recorded - signal are possible. Assessing imminence of a catastrophic failure with amplitude-frequency distribution of the recorded signal (as e.g., in \cite{Amitrano&al2005} for example) becomes invariably biased.

\subsection{Harnessing the attenuation phenomenon: Signal co-detection}
\label{codetect}

The attenuated wave amplitude may fall below a sensor detection threshold. Consider a small amplitude event, attenuation dictates that such an event will be detected only if it occurs close to a sensor (Fig. \ref{obs_attenuation}b). In contrast, for large initial amplitude, an event could be detected simultaneously by more than one sensor in the network (Fig. \ref{obs_attenuation}c).

Hence, event co-detection by multiple sensors occurs only if its initial amplitude of the event is sufficiently large. Naturally, co-detection would also depend on the sensor network geometry and detection threshold. Nevertheless, for any practical sensor network, signal co-detection offers a simple means for directly assessing the likelihood of a large amplitude event in real time (without the need of post-processing recorded signals). 
A simple on/off trigger (detection or no-detection) combined with an accurate time synchronization (to ensure that the same event is detected simultaneously) would thus be sufficient to infer the minimum amplitude of the source as well as an approximate location of the source (depending on the location of the sensors in the network that detect simultaneously the same event).

\section{A quantitative approach}
\subsection{Model description}
To quantitatively evaluate the effect of attenuation phenomenon on the detection of precursory acoustic activity and propose methodology for early warning system based on the co-detection method, a numerical model was used to systematically evaluate the potential of such an approach. Several models for heterogeneous material failure and fracturing exist in the literature (see \cite{Sornette2006,Alava&al2006,Bonamy2011} for a review).
We selected the Fiber Bundle Model (FBM) for its simplicity and generality as a useful framework for systematically studying processes preceding global failure \cite{Pierce1926,Daniels1945,Gomez&al1993,Kloster&al1997,Alava&al2006,Pradhan&al2010,Faillettaz&Or2015}. In essence, the model represents natural heterogeneous materials as a set of elasto-brittle fibers that are mechanically loaded in parallel (Fig. \ref{obs_attenuation}d), each fiber deforms in a linear elastic manner and breaks instantly at its prescribed rupture strength (whose values are drawn from a prescribed probability distribution). The load carried by the failed fiber is then redistributed according to specific rule, that could give rise to cascading failure events ("avalanches"). The FBM framework provides a simple and mechanically consistent means to study precursory signals preceding catastrophic rupture using recorded signals only. 
 Additionally, the discrete nature of failure events offers a direct link with acoustic emissions suggested for monitoring such progressive failures \cite{Michlmayr&al2012}. 
The application of 2-D spatially explicit FBM-based models (Fig \ref{obs_attenuation}d) have already provided new insights especially in term of early warning perspectives. Recent developments \citep{Faillettaz&Or2015} highlighted the important roles of spatial correlation of mechanical properties and load redistribution rules on failure statistics and global failure of the FBM.
The failure mode for the spatially explicit FBM varies dramatically with increasing spatial correlation length of mechanical properties and localized load sharing rules. Systems with similar composition of mechanical elements exhibit a dramatic transition from ductile and diffuse damage for global load sharing  (DFBM) to brittle single failure for correlated and local load sharing (LFBM). 
  These changes in mechanical responses also affect the statistical properties of fiber failure avalanche (micro-cracks) activity preceding rupture and sought after in various early warning scenarios \citep{Amitrano2012}. 
  
 Whereas diffuse damage behavior exhibits clear precursory signals (such as increased seismic activity prior to global failure), brittle failure occurs abruptly with only few precursors. Although increasing spatial correlations of mechanical properties promotes abrupt ruptures at lower external load, a "universal" global failure criterion based on macroscopic properties was obtained. This criterion is independent of the rupture mode, stress redistribution rules, or the spatial organization of mechanical properties.

However, the study of \cite{Faillettaz&Or2015} did not account for attenuation phenomena and the useful results in terms of early warning perspective might be biased under natural conditions where the signal recorded by a sensor at a given location has been attenuated. 
We thus aim to use this fiber bundle model as a tool to investigate the impact of attenuation on the existence of precursory signs of impeding rupture, especially signs linked to the statistical properties of the avalanche size frequency distribution prior to global failure.

\subsection{Spatial FBM with attenuation - model development}
We introduced signal attenuation effects into the spatial FBM developed by \cite{Faillettaz&Or2015}.
Attenuated amplitudes are computed assuming (i) that they are proportional to the initial "failure avalanche" size (i.e., the number $n$ of failed fibers at each load increment), and (ii) accounting for geometrical spreading effects only (the divergence of elastic energy density with radial distance from the source). In this simplified representation, we ignore several important interactions such as intrinsic and scattering attenuation that depend on material properties and frequency content and thus are difficult to quantify, in contrast with geometrical decay that occurs always. 
Knowing both the location of a sensor $X_{\rm sensor}$ and each single fiber failure $X_{ff}$ during the computed failure avalanche, the attenuated amplitude $A_a$ is simply evaluated by summing the individual attenuated amplitudes of each individual fiber failure reaching the sensor, i.e., 
\begin{equation}
\label{att}
A_{\rm a} = \sum_{ff=1}^n \frac{A_0}{\| X_{ff}-X_{\rm sensor} \|}
\end{equation}
where $A_{\rm a}$ is the attenuated amplitude reaching the sensor, $A_0$ the initial amplitude caused by a single fiber failure, $X_{ff}$ the location of the failed fiber and $X_{\rm sensor}$ the sensor location.
It is thus possible to access the statistical distribution of the attenuated avalanche sizes and their evolution prior global failure.

\subsection{Results}

\subsubsection{Size-frequency distributions for attenuated elastic signal}
\begin{figure*}

\noindent\includegraphics[width=\textwidth]{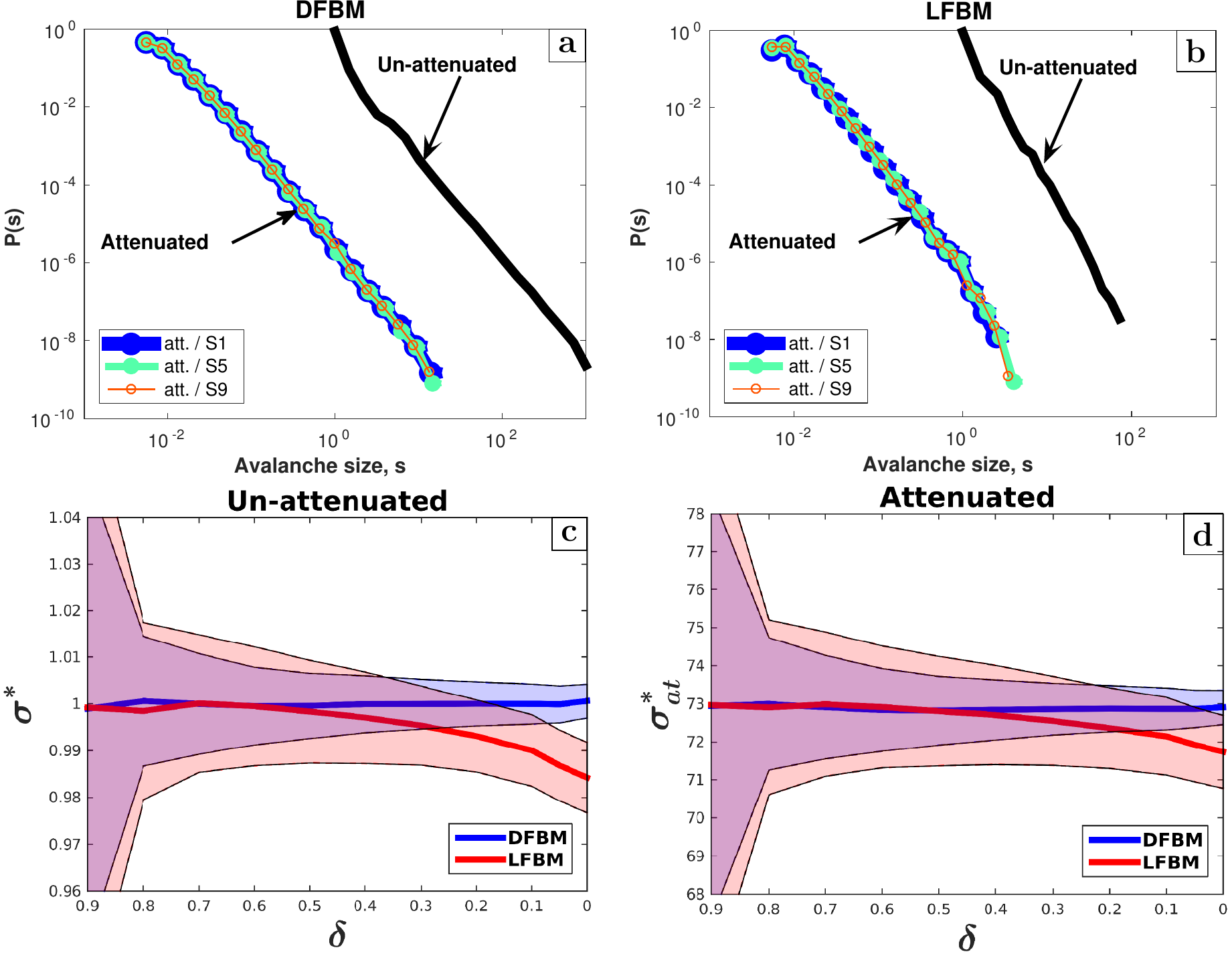}

\caption{\label{SFD} First line (a and b): Avalanche size frequency distribution from original (un-attenuated) data (in black) and attenuated data at three different random locations for DFBM (\textit{a}) and LFBM (\textit{b}). Second line: Evolution of rupture criterion (see text) defined in \cite{Faillettaz&Or2015} when approaching rupture for original (un-attenuated, \textit{c}) and attenuated (\textit{d}) data (with $\delta=\frac{\sigma_c-\sigma}{\sigma_c}$, $\sigma_c$ marking the stress at the critical point). }
\end{figure*}

Notwithstanding its simplicity, the model supports systematic evaluation of the effects of  attenuation on the statistical behavior of micro-crack activity prior rupture for different rupture mechanisms, i.e. brittle-like to ductile-like rupture.
\cite{Faillettaz&Or2015} have shown that load redistribution rules play a major role in the rupture behavior ranging from brittle-like rupture for local load redistribution (LFBM) to ductile-like behavior for global load redistributions (DFBM). 
Fig. \ref{SFD} shows a comparison between avalanche size frequency distribution (SFD) deduced from FBM failure events for the original (unattenuated) and attenuated events (taking $A_0 = 1$ in Eq. \ref{att} to permit comparison of the results) computed at different random location of sensors. 
Results indicate that the exact location of a sensor has no effect on the size frequency distribution of attenuated avalanches (Fig. \ref{SFD}), irrespective of the nature of global rupture behavior, i.e., brittle-like (LFBM) or ductile-like (DFBM). Statistical distribution of attenuated avalanches size remains indeed similar from one sensor location to another.
This unintuitive result is of practical interest as one would expect that sensor location would directly influence the recorded signal and thus affect predictive capabilities.

Moreover Fig. \ref{SFD} shows that SFD characteristics are qualitatively very different between original (unattenuated) and attenuated simulations. Avalanche sizes remain power law distributed when considering attenuation, but the associated $b$-exponent (exponent characterizing the power law exponent) decreases for the attenuated case, suggesting that the proportion of small avalanches  (group of fiber failures) increases in SFD deduced considering attenuation. 
%
%
The modified  statistical behavior of small recorded events with attenuation relative to the original behavior suggests that this information  might  not be appropriate for early-warning purposes. 

\subsubsection{Signal evolution towards global failure}
\label{towardsrupture}

\begin{figure*}

\noindent\includegraphics[width=\textwidth]{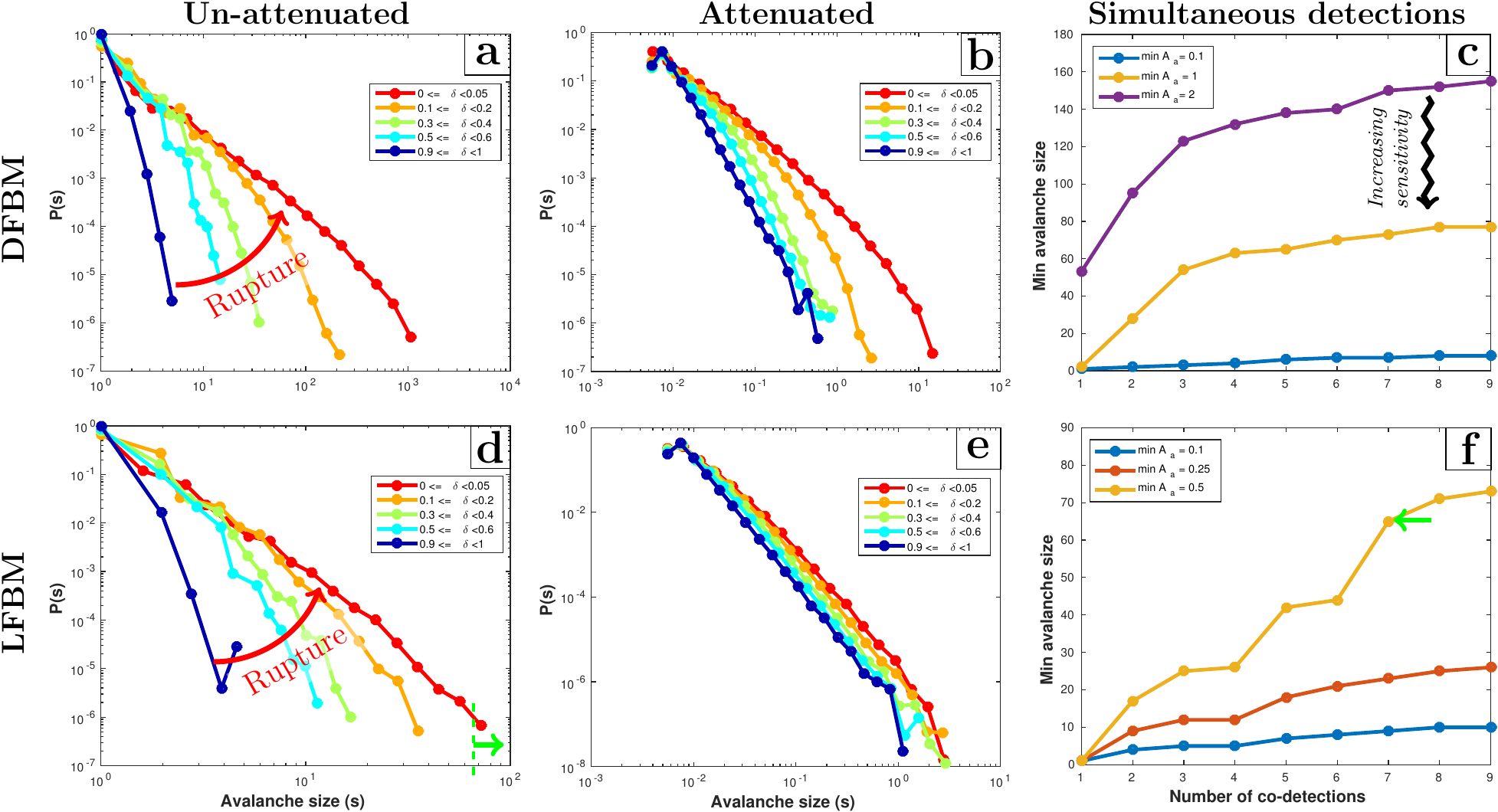}
\caption{\label{evol}Evolution of size frequency distribution when approaching rupture for original (un-attenuated, left, \textit{a},\textit{d}) and attenuated (middle,\textit{b},\textit{e}) avalanches in DFBM (top row, \textit{a},\textit{b}) and LFBM (bottom row, \textit{d},\textit{e}). Right column shows the minimum original (un-attenuated) avalanche size as a function of the number of sensors simultaneously detecting the same event for different detection thresholds (i.e., minimum $A_a$ been detected) for DFBM (\textit{c}) and LFBM  (\textit{f}): For example in LFBM, if an event is co-detected by 7 sensors with a threshold of 0.5 (green arrow in \textit{f}), the minimum avalanche size of the corresponding unattenuated event would be 70 indicating an imminent failure of the bundle ($\delta <0.05$, green arrow in \textit{d}). }
\end{figure*}

In order to characterize how fiber failure event size distribution evolves toward the macro-failure, we computed SFD for successive bins of the control parameter $\delta$. As the simulations were realized under stress controlled conditions, $\delta=\frac{\sigma_c-\sigma}{\sigma_c}$, $\sigma_c$ marking the stress at the critical point. $\delta$ represents also the relative stress towards failure (with a value of $\delta = 0$ marking failure).
To ensure reliable statistical representation we stacked 200 simulations for each configurations (DFBM and LFBM) taking a total number of fibers equal to 256$\times$256.

Fig. \ref{SFD} (right panel) shows the evolution towards rupture of a mechanical metric termed the  "damage weighted stress" $\sigma^{\star}$ proposed by \cite{Faillettaz&Or2015} that accounts for damage accumulation and expressed as
$\sigma^{\star}=\frac{\sigma^{\rm tot}}{E \; (1-D)D}$
 where $\sigma^{tot}$ is the total stress applied on the bundle and $D$ the associated damage defined as the ration between cumulative number of avalanches (i.e. the total number of failed fibers) and the total number of fibers in the bundle.
This "damage weighted stress" appears to be constant close to 1 for DFBM and slightly decreasing for LFBM when approaching global failure as noted by \cite{Faillettaz&Or2015}. 
Additionally we define damage in the presence of attenuation as the ratio between the cumulative attenuated avalanche size and the total number of fibers,i.e., an "attenuated damage weighted stress". Results show that evolution of both damage metrics approach global rupture in a very similar way.
This suggests that, although different, the analysis of attenuated signals provides useful information on the imminence of global rupture.

Fig. \ref{evol} depicts the evolution of size frequency distribution of original (unattenuated) and attenuated avalanche size close to global failure of DFBM and LFBM: 
 The SFD appears to change just prior to global failure, the simulated $b$-exponent increases as the system approaches catastrophic rupture in agreement with analytical results for ductile-like rupture (DFBM, \cite{Pradhan&al2005}). Increasing $b$-exponent prior rupture was suggested to constitute a useful tool for assessing the mechanical state  of a failure sensitive slope of geological structure \citep{Pradhan&al2005,Amitrano2012}.
Such effect is however not recovered very clearly when considering attenuated amplitudes (Fig. \ref{evol} middle panel), especially in the case of brittle-like rupture (LFBM), making this property less useful for early-warning purpose in practical real cases (with attenuation).
On the other hand, ability to detect the occurrence of large (un-attenuated) events would provide valuable information on the proximity of rupture as the larger events only occur near the global rupture and would thus provide a way to assess slope stability at proximity to rupture.

\subsubsection{Co-detection as a component of early warning}
The FBM model was used to evaluate the simple theoretical method based on co-detection proposed in section \ref{codetect}. As shown in Fig. \ref{evol} (a or b), the minimum initial amplitude of the micro-crack generated in the failing earth material or slope (represented by the spatial FBM) could be  assessed by investigating both the number sensors co-detecting an event and by tuning sensor  detection threshold. 
The larger the number of simultaneous detections (co-detection events), the larger the initial amplitude of the local failure event. Interestingly, higher sensor threshold values, select for co-detection of larger initial amplitude event. 
Fig. \ref{evol} shows that taking a sufficiently large threshold and number of co-detection events, the minimum size of the unattenuated events would fall in the range of $0<\delta<0.1$ i.e. just prior to rupture. For example in the LFBM, if an event is co-detected by 7 sensors with a threshold of 0.5 (yellow line, green arrow in Fig. \ref{evol}f), the minimum size of the unattenuated event would be 70 (interpreted as 70 fibers failing at once) indicating an imminent failure of the bundle ($\delta <0.05$, green arrow in Fig. \ref{evol}d).
 
These results suggest that the co-detection method could provide a simple and cost-effective way of detecting the occurrence of very large events that announce impeding global rupture, provided that sensor detection threshold is correctly tuned. In practice such detection thresholds could be tuned by posterior analysis introducing a flexible way to investigate the large event.
Co-detection method would thus constitute a simple, flexible and efficient way to assess imminence of rupture  in real time without processing a large amount of data. 

A perquisite for the implementation of the co-detection method is capability to accurately synchronize time stamps among sensors in the network.
The parameters for such time synchronization would vary with the material properties and the network geometry. For example, to detect the same event simultaneously in rock, the time synchronization should be better than a millisecond, taking 3 meters as the minimum distance between two sensors in the network and an average elastic wave speed of 3000 $\rm m.s^{-1}$ in rock \citep{Hamilton1978}.

Note that the co-detection method is general and should hold for any waves propagating in any material considering homogeneous attenuation properties. 
However, we do not address in this study neither intrinsic attenuation nor scattering effects  (that highly depend on frequency content of the wave). Both effects further enhance attenuation phenomenon, leading to drastic attenuation especially at high frequency. 
The co-detection method remains valid as long as the attenuated wave is detectable, i.e., amplitude higher than the background noise.
In practice the geometry of the sensor network has to be adapted to the studied frequency domain: The higher the frequency range, the smaller distance between sensors in the network. In several field tests it was found that damping of acoustic signals requires a dense network of acoustic emission sensors leading to strong practical limitations \citep{Dixon&al2015}.
New emergent techniques such as distributed fiber-optic acoustic emission acquisition could also significantly extend the effective coverage with the ability to detect elastic waves continuously along the optical cable \citep{Michlmayr&al2015}.
Alternatively, signals at lower frequencies could offer observability advantage provided they are linked to precursory mechanical failure events. Subjected to less attenuation, such waves can be detectable at a larger distance, with seismic sensors or accelerometers.

\section{Application}

To test the applicability of the proposed co-detection method, we used cold laboratory experiments that recorded acoustic emissions activity generated during a force-controlled loading of layered snow samples containing a weak layer \citep{Reiweger&al2015}. The analysis of the acoustic data revealed a decrease of the $\beta$-exponent of the complementary cumulative distribution of event energy during the catastrophic failure, as evidenced the "numerical" behavior depicted in Section \ref{towardsrupture} and Fig. \ref{evol}.
Six acoustic sensors located in the snow sample were recording simultaneously the acoustic activity prior to the catastrophic failure, enabling thus the co-detection to be tested. We show here results for loading experiment on sample TRA 5 as such small sample experienced a clear catastrophic failure with a clear drop in $\beta$-exponent prior to failure. 

\begin{figure*}
\noindent\includegraphics[width=\textwidth]{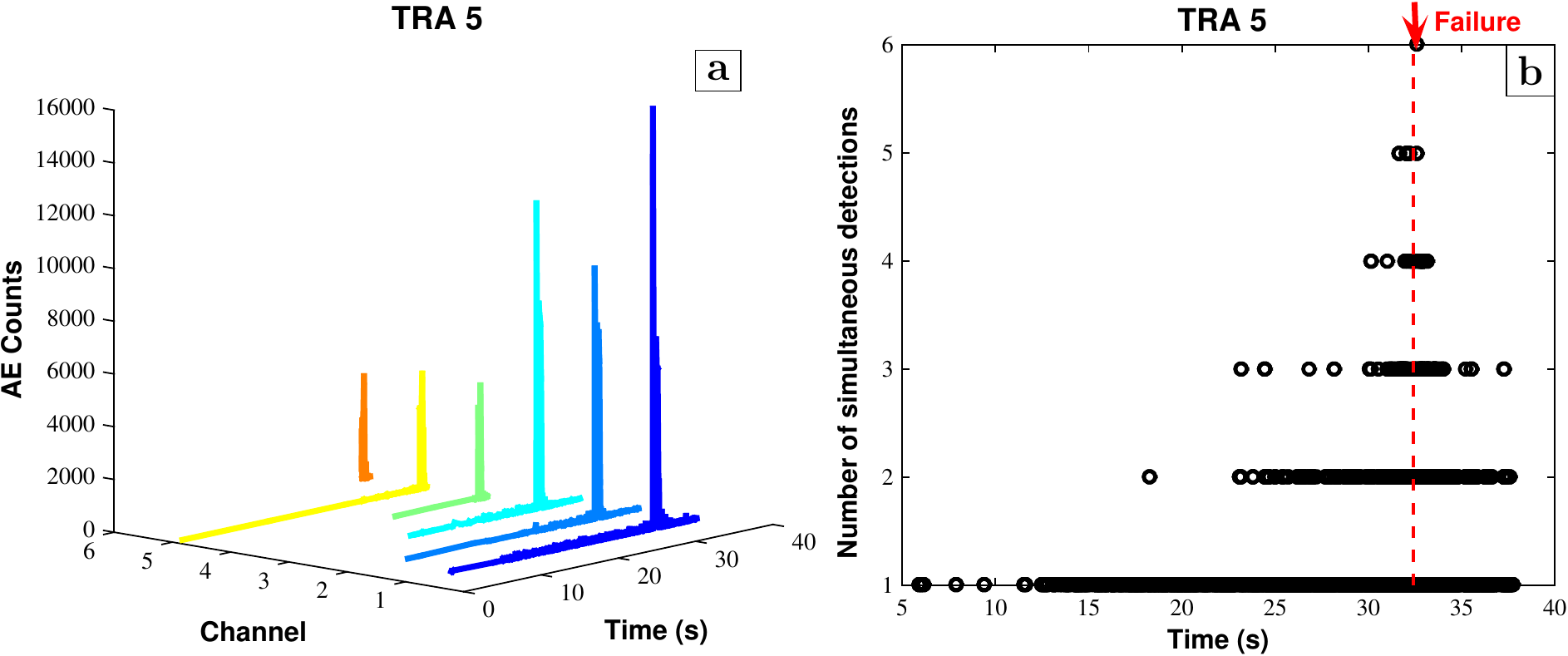}
\caption{\label{ingrid} Experimental data of recorded acoustic emission (AE) activity during a force-controlled loading of layered snow sample \citep{Reiweger&al2015}
(a) Recorded acoustic emissions (AE) counts from the six sensors (experiment "TRA5"), (b) Number of sensors detecting simultaneously the same event. Red dashed line marks the failure of the snow sample. }
\end{figure*}

Results show a progressive increase in the number of AE event co-detections prior to the catastrophic failure of the sample, with a peak of six co-detection events (all sensors) at the failure (Fig. \ref{ingrid}).
Such a clear increasing trend in the number of co-detections is an encouraging result which confirms a possible application to early warning. In such an example the catastrophic failure could have been predicted around one minute in advance, when 5 simultaneous co-detections occurred.

\section{Summary and conclusions}

Early warning systems are often based on the monitoring of the temporal evolution of external parameters such as geometry, surface displacements or meteorological variable (e.g., rainfall duration and intensity).
Our study proposes a method based on continuous monitoring and interpretation of acoustic emissions and their characteristics as surrogates of mechanical states of the system (hillslope, glacier, snow cover). The method capitalizes on both heterogeneity and attenuation properties of natural media to develop a new strategy for early warning systems. The study introduces a heuristic and simple method based on co-detection of elastic waves traveling through natural media.
It requires a network of (seismic/acoustic) sensors on a potential unstable slope and monitor events detection in real time. Real time processing of measured events that are detected simultaneously on more than one sensor (co-detected) would then enable to easily access to their initial size as well as their initial location. Such method provides a simple means to access characteristics and temporal evolution of surrogate variables linked to hillslope damage and mechanical state.  
Simple numerical model based on Fiber Bundle Model accounting for geometric attenuation confirms the early warning potential of co-detection. Results suggest that although statistical properties of attenuated signal amplitude could lead to misleading results, monitoring the emergence of large events announcing impeding failure is possible even with attenuated signals depending on sensor network geometry and detection threshold.
In this context, the sensors must not be very sensitive (i.e., a low threshold is not needed) to assess slope stability but the network need to be precisely synchronized: Temporal synchronization between sensors must be sufficiently accurate to reliably classify events detected simultaneously by multiple sensors.
Preliminary application of the proposed method to acoustic emissions during failure of snow samples have confirmed the potential usefulness of co-detection as indicator for imminent failure.  More tests are needed to refine aspects of threshold-co-detection and their statistical properties for real early warning systems.

\textbf{Acknowledgments.}

This study was supported in part by the X-Sense2 project of the Swiss National Science Foundation with funding by Nano-Tera.ch from the Swiss Confederation, and by the TRAMM project "Triggering of Rapid Mass Movements" funded by the Competence Center Environment and Sustainability (CCES) of the ETH domain (Switzerland).

\bibliographystyle{agufull08}
\bibliography{/home/jfaillet/boulot_3G/Article/my_bib/shorttitles,/home/jfaillet/boulot_3G/Article/my_bib/my_bib}

\end{document}